\documentclass[12pt]{article}
\usepackage[reqno]{amsmath}
\usepackage{amsthm}
\usepackage[dvips]{color,graphicx}
\usepackage{amssymb} 
\def\href#1#2{#2}
\def\IP{\relax{\rm I\kern-.18em P}}

\def\s{{\sigma}}
\def\cN{{\tilde N}}
\def\Mat{{\rm Mat}}
\def\B{{\sf b}}
\newcommand{\lcm}{\text{\normalfont lcm}}
\newtheorem{lem}{Lemma}
\newcommand{\nc}{\newcommand}
\newcommand{\beq}{\begin{equation}}
\newcommand{\eeq}{\end{equation}}
\newcommand{\beqa}{\begin{eqnarray}}
\newcommand{\eeqa}{\end{eqnarray}}

\def\om{\omega}
\def\bfn{n}

\def\bT{{\bar T}}
\def\sg{{\cal G}} 
\def\asg{{\widehat{\sg}}}
\def\bJ{{\bar J}}
\def\bz{{\bar z}}
\def\pl{\partial}
\def\bpl{\bar \partial} 
\def\Ad{{\rm Ad}}
 
\def\st{{\cal T}}
\def\dim{{\rm dim\,}}
\def\rank{{\rm rank\,}}
\def\id{{\rm id}}
\def\c{\gamma} 
\def\C{{\cal C}} 
 
\def\cA{{\cal A}}
\def\cH{{\cal H}}
\def\cJ{{\cal J}}
\def\bJ{{\bar J}} 

\nc{\nn}{\nonumber}
\def\Tor{{\rm Tor}} 

\def\o{\otimes}

\def\a{\alpha}
\def\b{\beta} 
\def\IP{\relax{\rm I\kern-.18em P}}
\def\tr{{\rm tr\ }}

\def\QC{\mathbb{C}}

\def\QZ{\mathbb{Z}}

\def\bZ{\QZ}
\def\U{{U}}

\def\nn{\nonumber}

\def\cH{{\cal H}}

\def\a{\alpha}

\input epsf
\newcount\figno
\def\fig#1#2#3{
\par\begingroup\parindent=0pt\leftskip=1cm\rightskip=1cm\parindent=0pt
\baselineskip=11pt
\global\advance\figno by 1
\epsfxsize=#3
\centerline{\epsfbox{#2}}
\vskip 12pt
{\bf Figure \the\figno:} #1\par
\endgroup\par
}
\def\figlabel#1{\xdef#1{\the\figno 
\mbox{ }}}
\def\encadremath#1{\vbox{\hrule\hbox{\vrule\kern8pt\vbox{\kern8pt
\hbox{$\displaystyle #1$}\kern8pt}
\kern8pt\vrule}\hrule}}

\begin{document}
\baselineskip=17pt
\title{\bf Branes on Group Manifolds,\\[2mm] Gluon Condensates, \\[2mm] 
                     and twisted K-theory \\[5mm] }
\author {{\sc Stefan Fredenhagen and Volker Schomerus} \\[2mm]
                  Albert-Einstein-Institut
      \\ Am M{\"u}hlenberg 1, D--14476 Golm, Germany \\[1mm] }
\vskip.2cm
\date{December 19, 2000}
%
\begin{titlepage}      \maketitle       \thispagestyle{empty}

\vskip1cm
\begin{abstract}
\noindent  
In this work we study the dynamics of branes on group 
manifolds $G$ deep in the stringy regime. After giving a 
brief overview of the various branes that can be 
constructed within the boundary conformal field 
theory approach, we analyze in detail the condensation 
processes that occur on stacks of such branes. At 
large volume our discussion is based on certain 
effective gauge theories on non-commutative `fuzzy' 
spaces. Using the `absorption of the boundary spin'-%
principle which was formulated by Affleck and Ludwig 
in their work on the Kondo model, we extrapolate the 
brane dynamics into the stringy regime. For 
supersymmetric theories, the resulting condensation 
processes turn out to be consistent with the existence 
of certain conserved charges taking values in some 
non-trivial discrete abelian groups. We obtain strong 
constraints on these charge groups for $G = SU(N)$. 
The results may be compared with a recent proposal 
of Bouwknegt and Mathai according to which charge
groups on curved spaces $X$ (with a non-vanishing 
NSNS 3-form field strength $H$) are given by 
the twisted K-groups $K_H^*(X)$.    
\end{abstract}
\vspace*{-18.9cm}
{\tt {AEI-2000-079 \hfill hep-th/0012164 }}\\
{\tt \phantom{{UUITP-nn/99} \hfill hep-th/0012164}}
\bigskip\vfill
\noindent
\phantom{wwwx}{\small e-mail:}{\small\tt stefan@aei-potsdam.mpg.de; 
vschomer@aei-potsdam.mpg.de} 
\end{titlepage}

\section{Introduction}

During the last years, the study of branes and their 
dynamics has lead to many new insights into string 
and M-theory. Much of this study was done in the 
large volume regime where geometric techniques 
provide reliable information. The extrapolation 
into the stringy regime usually requires new 
methods from boundary conformal field theory, 
in particular when the bulk supersymmetry is 
not maximal. The analysis of strings and branes 
on group manifolds gives us a good handle on such
issues. The large symmetry of group manifolds $G$
makes string  theory on $G$ rather tractable while, 
on the other hand, group manifolds display many 
interesting new features that do not appear in 
flat spaces. Most importantly, their non-vanishing 
curvature along with the string equations of motion
imply that they carry a non-vanishing NSNS 3-form 
field strength $H$. Moreover, models of strings
and branes on group manifolds are used as a starting 
point in perturbative string constructions for many 
other backgrounds. 
\bigskip   

Our focus in this work is on bound state formation
of branes and on finding appropriate conserved 
quantities (charges) that encode the essential
features of the brane dynamics. D-branes in a 
background $X$ may be characterized by their ability 
to carry RR-charges \cite{Pol}. The latter are assigned to 
arbitrary configurations of branes, stable and 
unstable, and they are conserved during all dynamical 
processes. In the world-sheet description,  D-brane 
configurations correspond to boundary conditions for 
some 2D conformal field theory and their condensation 
is induced by relevant (or marginally relevant) boundary 
operators. The infra-red (IR) fixed point of the associated 
renormalization group (RG) trajectory provides the 
world-sheet theory for the decay product that is
reached after the condensation has occurred. In this
framework, the conserved RR-charges are simply 
RG-invariants.
\medskip

By construction, the brane charges take values in 
some discrete abelian group. Obviously, the latter 
contains a lot of information about the brane 
dynamics (i.e.\ RG trajectories) and hence it 
is rather difficult to find. On the other hand, 
there are many constructions in mathematics that 
assign discrete abelian groups to a background 
geometry $X$. These include the de Rham cohomology 
and various different K-theories. Very naively one 
might think that RR-charges take values in de Rham 
cohomology groups $H^*(X,\QZ)$ as they are associated 
with the n-form fields of super-gravity theories. It 
is by now well known that this naive expectation is 
incorrect  and that K-groups provide a much more 
realistic candidate for the group of RR-charges 
(see e.g.\ \cite{MiMo} and the more recent 
developments \cite{Wit1,Hor,FreWit,MooWit,Dia,Wit2} 
that were initiated mainly by \cite{Sen}).     
\smallskip

There exist various different K-theories that one
uses depending on the string theory under 
consideration. For type IIA/B theory in a 
background $X$, the relevant groups are given 
by the usual $K^*(X)$, provided that $X$ carries
a vanishing NSNS 3-form $H$. In dealing with the 
general case $H \in H^3(X,\QZ)$, Bouwknegt and 
Mathai proposed to employ the twisted K-groups 
$K_H^*(X)$. The latter are defined as K-groups of 
an algebra whose elements are sections of some 
bundle on $X$ taking values in compact operators. 
Morita invariance of algebraic K-theory implies 
that one recovers $K(X)$ for $H = 0$. When the 
$H$-field is torsion class, i.e.\ some integer 
multiple of it vanishes in $H^3(X,\QZ)$, the 
proposal of Bouwknegt and Mathai boils down to 
K-groups suggested in \cite{Wit1} (see also 
\cite{Kap} for an extensive discussion). 
\medskip
  
Having all these different groups at our disposal, 
it is important to decide which one gives the 
right answer, i.e.\ leads to some RG-invariants
in boundary conformal field theories. To begin 
with, this requires some background $X$ for which 
the various groups are actually different. In finding
such examples one needs to overcome de Rham's
theorem which claims that the non-torsion parts
of de-Rham cohomology and of the usual K-theory 
are isomorphic. \footnote{A finitely generated 
abelian group $C$ is of the form $C = \QZ  \oplus
\dots \oplus \QZ \oplus \QZ_{x_1} \oplus \dots \oplus 
\QZ_{x_n}$. The subgroup $\Tor(C) := \QZ_{x_1} \oplus 
\dots \oplus \QZ_{x_n} \subset C$ is called the {\em 
torsion} of $C$.}Hence, one has to work on string 
backgrounds for which the cohomology  and K-groups 
possess torsion parts. Such examples are known (see 
e.g.\ \cite{Bra}), but to study brane dynamics in such 
backgrounds still presents a challenge. 
\medskip

String theory on group manifolds provide an interesting 
class of examples for which the appropriate twisted K-theory 
is rather non-trivial while at the same time there exist
powerful methods to study the dynamics of branes. Group 
manifolds $G$ are curved so that by the string equations 
of motion they come equipped with a non-trivial $H$-field, 
known as the WZW 3-form. The latter is non-torsion and 
hence we expect the brane charges to take values in the 
twisted K-groups $K_H^*(G)$.
\smallskip 

The simplest example is given by $G = SU(2)$. It was shown 
in \cite{AlSc1} that branes on $SU(2)$ can wrap integer 
conjugacy classes. Their stability was also analyzed in 
\cite{BaDoSc,Paw}. Generically, conjugacy classes of $SU(2)$ 
are 2-spheres but there are two exceptional point-like classes 
provided by $\pm e$ where $e$ is the group unit. The large volume
analysis of \cite{AlReSc2} shows that an arbitrary spherical 
brane on $SU(2)$ may be obtained as a bound state of some 
sufficiently large stack of point-like branes, similarly to 
the effect described in \cite{Mye}. This implies 
that all branes carry a certain integer multiple of the 
charge of a single point-like brane. When the volume of 
$SU(2) \cong S^3$ is finite, the condensation can lead us 
from a sufficiently large stack of point-like branes at $e$ 
to a single point-like brane at $-e$ \cite{AlSc2}. Hence, 
the authors of \cite{AlSc2} concluded that the charge of a 
point-like brane does not take values in the integers but 
in some finite quotient thereof. We shall see that the result 
is in perfect agreement with the K-theoretic prediction. 
The investigations in \cite{AlSc2} were motivated by 
\cite{Tay} (see also \cite{Sta} and the more recent work
\cite{FigSta}). 
\smallskip
    
Other groups admit branes which wrap more general 
`twisted' conjugacy classes \cite{FFFS}. Their dynamics 
can still be analyzed even deep in the stringy regime and 
the analysis gives strong constraints on the possible 
charge groups. In particular, we shall see that they 
are all {\em finite} discrete abelian groups but our 
information will be much more detailed. For $G =
SU(N)$ we shall show that the charge group is of 
the form 
\begin{equation}  \label{mres} 
  C(SU(N),K) \ = \ \QZ_x \oplus \bigoplus_{\nu = 1}^{s} 
  \QZ_{x_\nu}
\end{equation} 
where $x = K/\gcd(K,\lcm(1, \dots, N))$ and the $x_\nu$ 
are known to divide $x$ when $N \leq 5$ or $N$ odd. For
even $N \geq 6$ our results on $x_\nu$ are slightly 
weaker and we defer their precise formulation to 
Section 4.3 below. The integer $K$ is determined by the 
NSNS 3-form $H \in H^3(SU(N),\QZ) \cong \QZ$. Branes 
wrapping ordinary conjugacy classes contribute the first 
summand to eq.\ (\ref{mres}) while the others come with 
twisted branes. Since the latter are more difficult to 
study, our information on the charges of twisted branes 
is not complete. For $G = SU(2)$ we shall show that $s=0$. 
In case of $G=SU(3)$, the comparison with twisted K-theory 
suggests that $s = 1$ and $x_1=x$.    
\bigskip

The plan of this paper is as follows. In the next section 
we will review the theory of maximally symmetric branes on 
group manifolds. In particular, we shall provide a complete
list of such branes and their associated open string 
spectra. Section 3 is devoted to the dynamics of branes
on group manifolds. We start with a brief summary and 
generalization of the results obtained in \cite{AlReSc2} 
for the large volume regime. The main aim of the section 
is then to explain how condensation processes can be 
studied deep in the stringy regime and to present 
explicit results on bound states. This information 
is then used in Section 4 to derive strong constraints
on the charge groups. In Section 5, finally, we make
some remarks on the comparison with twisted K-theory. 
These will be rather preliminary, though, because there
is not much known about the twisted K-groups beyond the
examples of $G = SU(2), SU(3)$. 

\section{D-branes on group manifolds} 

This section is devoted to the description of maximally 
symmetric branes on group manifolds. Following \cite{AlSc1,
FFFS}, we will begin with a brief review of their classical 
geometry. Then, in the second subsection, we shall present 
some basic results on the boundary conformal field theory 
of such branes.    

\subsection{The geometry of branes on group manifolds}

Strings on the group manifold of a simple and simply connected 
group $G$ are described by the WZW-model. Its action is evaluated on 
fields $g: \Sigma \mapsto G$ taking values in $G$ and it involves 
one (integer) coupling constant $k$, which is known as the `level'. 
For our purposes it is most convenient to think of $k$ as controlling 
the size (in string units) of the background. Large values of $k$ 
correspond to a large volume of the group manifold. When dealing 
with open strings at tree level, the 2-dimensional world sheet 
$\Sigma$ is taken to be the upper half plane $\Sigma = \{ z \in 
\QC | \Im z \geq 0\}$. 
\smallskip

Along the boundary of this world sheet we need to impose some 
boundary condition. Here we shall analyze boundary conditions 
that preserve the full bulk symmetry of the model, i.e.\ the 
affine algebra $\asg_k$. These boundary conditions are formulated 
in terms of the chiral currents 
$$ J (z) \ = \ k \, g^{-1}(z,\bz) \pl g(z,\bz ) \ \ \  , \ \ \ 
   \bJ (\bz) \ = \ -k \, \bpl g(z,\bz ) \, g^{-1}(z,\bz)\ \ . $$ 
Note that $J$ and $\bJ$ take values in the finite dimensional 
Lie algebra $\sg$ of the group $G$. Along the real line we glue 
the holomorphic and the anti-holomorphic currents  according to 
\begin{equation}
\label{glue}  
J(z) \ = \ \Lambda \bJ(\bz)  \ \ \ 
\mbox{ for all } \ \ \ z = \bz 
\end{equation}
where $\Lambda$ is an appropriate automorphism of the current  
algebra $\asg_k$ (see e.g.\ \cite{ReSc1}). The choice of 
$\Lambda$ is restricted by the requirement of conformal 
invariance which means that $T(z) = \bT(\bz)$ all along the 
boundary. Here $T, \bT$ are the non-vanishing components
of the stress energy tensor. They can be obtained through 
the Sugawara construction, as usual.%
\medskip   

The allowed automorphisms $\Lambda$ of the affine Lie algebra $\asg$
are easily classified. They are all of the form 
\begin{equation} \label{Lam} \Lambda \ = \ \Omega \circ \Ad_g \ \ 
       \mbox{ for some } \  g \in G \ \ . 
\end{equation}
Here, $\Ad_g$ denotes the adjoint action of the group element 
$g$ on the current algebra $\asg_k$. It is induced in the 
obvious way from the adjoint action of $G$ on the finite 
dimensional Lie algebra $\sg$. The automorphism $\Omega$ 
does not come from conjugation with some element $g$. More 
precisely, it is an outer automorphism of the current 
algebra. Such outer automorphisms $\Omega = \Omega_\omega$ 
come with symmetries $\omega$ of the Dynkin diagram of the 
finite dimensional Lie algebra $\sg$. One may show that 
the choice of $\omega$ and $g \in G$ in eq.\ (\ref{Lam}) 
exhausts all possibilities for the gluing automorphism
$\Lambda$ (see e.g.\ \cite{Kac}). 

Throughout this text, the groups $G = SU(N)$ will serve as 
our main examples. Their Dynkin diagrams $A_{N-1}$ possess 
the trivial symmetry $\omega = \id$ and one non-trivial 
involution $\omega$ when $N > 2$. We will not need explicit 
expressions for the associated maps $\Omega_\omega$, but the 
interested reader can find formulas e.g.\ in \cite{BiFuSc}.    
\bigskip

So far, our discussion of the possible types of gluing 
automorphisms $\Lambda$ has been fairly abstract. But it 
is possible to associate some concrete geometry with 
each choice of $\Lambda$. This was initiated in 
\cite{AlSc1} for $\omega = \id$ and extended to non-trivial
symmetries $\omega \neq \id$ in \cite{FFFS} (see also 
\cite{Gaw}, \cite{Stan}). 
\smallskip

\def\U{{\cal U}}
Let us assume first that the element $g$ in eq.\ (\ref{Lam}) 
coincides with the group unit $g = e$. This means that $\Lambda
= \Omega = \Omega_\omega$ is determined by $\omega$ alone. The
diagram symmetry $\omega$ induces an (outer) automorphism 
$\omega_\sg$ of the finite dimensional Lie algebra $\sg$ through 
the unique correspondence between vertices of the Dynkin diagram 
and simple roots. After exponentiation, $\omega_\sg$ furnishes
an automorphism $\omega_G$ of the group $G$. One can show that 
the gluing conditions (\ref{glue}) force the string ends to 
stay on one of the following $\omega$-twisted conjugacy 
classes
$$ C^\omega_u \ := \ \{\,  h u\,  \omega_G(h^{-1})\ |\ h \in G 
\, \} \ \ . $$
The subsets $C^\omega_u \subset G$ are parametrized 
by equivalence classes of group elements $u$ where the 
equivalence relation between two elements $u,v \in G$ is
given by: $u \sim_\omega v$ iff $v \in C^\omega_u$. Note 
that this parameter space $\U^\omega$ of equivalence classes
is not a manifold, i.e.\ it contains singular points.  
\smallskip

To describe the topology of $C^\omega_u$ and the parameter 
space $\U^\omega$ (at least locally), we need some more 
notation. By construction, the action of $\omega_\sg$ on 
$\sg$ can be restricted to an action on the Cartan subalgebra $\st$. 
We shall denote the subspace of elements which are invariant 
under the action of $\omega_\sg$ by $\st^\omega \subset \st$. 
Elements in $\st^\omega$ generate a torus $T^\omega \subset G$. 
One may show that the generic $\omega$-twisted conjugacy class
$\C^\omega_u$ looks like the quotient $G/T^\omega$. Hence, 
the dimension of the generic submanifolds $\C^\omega_u$ 
is $\dim G - \dim T^\omega$ and the parameter space has 
dimension $\dim \st^\omega$ at all but finitely many 
points. In other words, there are $\dim \st^\omega$ 
directions transverse to a generic twisted conjugacy 
class. This implies that the branes associated with the 
trivial diagram automorphism $\omega = \id$ have the largest 
number of transverse directions. It is given by the rank of 
the Lie algebra. 

As we shall see below, not all these submanifolds $C^\omega_u$ 
can be wrapped by branes on group manifolds. There exists some 
integrality requirement that can be understood in various ways, 
e.g.\ as quantization condition within a semiclassical analysis 
\cite{AlSc1} of the brane's stability 
\cite{BaDoSc,Paw}. This implies that there is only a finite set of 
allowed branes (if $k$ is finite). The number of branes depends 
on the volume of the group measured in string units. 
\smallskip

Let us become somewhat more explicit for $G = SU(N)$. The 
simplest case is certainly $N=2$ because there exists no 
non-trivial diagram automorphism $\omega$. The conjugacy 
classes $C^\id_u$ are 2-spheres $S^2 \subset S^3 \cong 
SU(2)$ for generic points $u$ and they consist of a 
single point when $u = \pm e$ in the center of $SU(2)$.  
More generally, the formulas $\dim SU(N) = (N-1)(N+1)$ and 
$\rank SU(N) = (N-1)$ show that the generic submanifolds 
$C^\id_u$ have dimension $\dim C^\id_u = (N-1)N$. In 
addition, there are $N$ singular cases associated with 
elements $u$ in the center $\bZ_N \subset SU(N)$. The 
corresponding submanifolds $C^\id_u$ are 0-dimensional. 
Note that all the submanifolds $C^\id_u$ are even 
dimensional. Similarly, the generic manifolds $C^\omega_u$ 
for the non-trivial diagram symmetry $\omega$ have 
dimension $\dim C^\omega_u = (N-1)(N+1/2)$ for odd N 
and $\dim C^\omega_u = N^{2}-N/2-1$ whenever 
$N$ is even. For some exceptional values of 
$u$, the dimension can be lower.        
\smallskip

So far we restricted ourselves to $\Lambda = \Omega_\omega$ being 
a diagram automorphism. As we stated before, the general case 
is obtained by admitting an additional inner automorphism of 
the form $\Ad_g$. Geometrically, the latter corresponds to rigid 
translations on the group induced from the left action of $g$ 
on the group manifold (see e.g.\ \cite{ReSc2}). The freedom of 
translating branes on $G$ does not lead to any new charges and 
we shall not consider it any further, i.e.\ we shall assume 
$g = e$ in what follows. 

\subsection{Conformal field theory} 

The branes we considered in the previous subsection may be described
through an exactly solvable conformal field theory. In particular, 
there exists a complete understanding of the open string spectra
based on the work of Cardy \cite{Car} and of Birke, Fuchs and 
Schweigert \cite{BiFuSc}. 
\smallskip

We shall use $\Xi = (\alpha,\omega)$ to label the boundary conformal 
field theories. The label $\alpha$ is taken from some index set 
$\cJ^\omega_k$ depending on the choice of the diagram automorphism 
$\omega$ and on the level $k$. For the trivial diagram automorphism 
$\omega = \id$, the set $\cJ_k = \cJ^\id_k$ coincides with the  
set of primaries of the affine Kac-Moody algebra $\asg_k$. As is well 
known, $\cJ_k$ is a certain subset of the set $\cJ = \cJ_\infty$ of 
equivalence classes of irreducible representations for the finite 
dimensional Lie algebra $\sg$. The automorphism $\omega$ generates 
a map $\varpi_k : \cJ_k \rightarrow \cJ_k$. In fact, given an 
irreducible representation $\tau$ of $\sg$, we can define another 
representation by composition $\tau \circ \omega_\sg$. The class 
of $\tau \circ \omega_\sg$ is independent of the choice of $\tau \in 
[\tau]$ and so we obtain a map $\varpi: \cJ \rightarrow \cJ$. The 
latter descends to $\cJ_k \subset \cJ$. A label $i \in \cJ_k$ is 
said to be ($\omega$-)symmetric, if it is invariant under the 
action of $\varpi$, i.e.\ if $\varpi i = i$. The subset of 
symmetric labels will be denoted by $\cJ_k^\omega \subset \cJ_k$. 
According to the results of \cite{Car,BiFuSc}, the labels $\alpha$ 
for branes associated with the diagram automorphism $\omega$ take 
values in the set $\cJ^\omega_k$. 
\smallskip

These very formal constructions can be understood as follows: 
obviously we would like to think of $\alpha$ in $\Xi = (\alpha,\omega)$ 
as labeling the position of the brane transverse to the $\omega$-twisted 
conjugacy classes. As we explained before, the transverse space is locally
given by $\st^\omega$. This fits nicely with our description of the 
sets $\cJ^\omega_k$. In fact, by construction the labels $\alpha
\in \cJ^\omega_k$ run through a set of points on some lattice of 
dimension $\dim \st^\omega$. When $\omega = \id$, this lattice 
coincides with the weight lattice of $\sg$. 
\medskip
  
Our main goal here is to explain the open string spectra that come
with these branes. For a pair of boundary labels $(\a,\omega), 
(\beta,\omega)$ associated with the same diagram automorphism $\omega$, 
the partition function is of the form  
\begin{equation} \label{part} Z^\omega_{\a \b}(q) \ 
     = \ \sum_{j \in \cJ_k} {\bfn}_{j \a}^
          {\omega; \b} \chi_j (q) \ \ . 
\end{equation} 
Here, $\chi_j(q)$ denote the characters of the current algebra 
$\asg_k$ and $\a,\b \in \cJ^\om_k$. Consistency requires the numbers 
${\bfn}_{j \a}^{\omega; \b}$ to be non-negative integers. 

There exists a very simply argument due to Behrend et al.\ \cite{BPPZ1} 
which shows that the numbers  ${\bfn}_{j \a}^{\om ; \b}$ give rise to a 
representation of the fusion algebra of $\asg_k$. This means that they 
obey the relations 
\begin{equation}
 \label{nfus}
  \sum_{\b \in \cJ^\om_k} {\bfn}_{i \a}^{\om; \b} \ 
    {\bfn}_{j \b}^{\om; \c}  \ = \ 
    \sum_{k \in \cJ_k} N_{ij}^{\ k} {\bfn}_{k \a}^{\om ; \c}\ \ , 
\end{equation}
where $N_{ij}^k$ are the fusion rules of the current algebra $\asg_k$. 
The argument of \cite{BPPZ1} starts from a general ansatz for the boundary 
state assigned to $(\a,\omega)$. Using world sheet duality, one can express 
the numbers ${\bfn}$ in terms of the coefficients of the boundary states
and the modular matrix $S$ for the current algebra $\asg_k$. The general form 
of this expressions is then sufficient to check the relations (\ref{nfus})
(see \cite{BPPZ1} for details). 
 
An explicit construction for the numbers $n$ is given in \cite{FuScI,BiFuSc}. 
Let $\omega$ be a diagram automorphism, as before. Then the numbers $n^\om$
are of the form    
\begin{equation}
\label{annuluscf}
\bfn^{\omega;\beta}_{i \alpha} \ = \ 
\sum_{\lambda \in \cJ^\omega_k}
\frac{S^{\omega \,*}_{\lambda\beta}S^{\omega}_{\lambda \alpha}
S_{\lambda i }}{S_{\lambda\, 0}} \ \ \ \mbox{ for } \ \ \  
\a,\b \in \cJ^\om_k \ \ \mbox{ and } \ \ i \in \cJ_k\ \ .  
\end{equation}
The matrix $S^{\omega}$ is a unitary matrix with matrix elements 
$S^\om_{\lambda \a}$ indexed by the $\varpi$-symmetric labels 
$\lambda, \alpha \in \cJ^\om_k$, i.e.\ they obey $\varpi \lambda = 
\lambda$ and $\varpi \a = \a$. When $\om = \id$, the matrix $S_\om$
coincides with the usual $S$-matrix so that Verlinde's formula 
implies 
$$ n^{\id; \b}_{i \a} \ = \ N^{\ \b}_{i \a} \ \ \ \mbox{ for all }
   \ \ \a,\b,i  \in  \cJ_k\ \ .$$
This reproduces Cardy's results on the boundary partition
functions \cite{Car}. For non-trivial automorphism $\om$, 
the matrix $S^\om$ describes modular transformations of 
twisted characters. Explicit formulas for $S^\om$ exist
(see e.g.\ \cite{BiFuSc}), but we will not need them here.  
\smallskip

Some aspects of the formulas (\ref{part}, \ref{annuluscf}) 
with $\a = \b$ can be understood geometrically. Let us note 
first that the partition functions of all our boundary theories 
are obtained by summing characters of the $\asg_k$ algebra. This 
reflects the fact that all the (twisted) conjugacy classes admit 
an obvious action of the Lie group $G$ by (twisted) conjugation. 

The spectrum of ordinary conjugacy classes can be explained 
in much more detail. For simplicity, we shall restrict to 
$G = SU(2)$. In this case, generic conjugacy classes are 
2-spheres and the space of functions thereon is spanned 
by spherical harmonics $Y^{j/2}_m, |m| \leq j/2$ and $j = 
0,2,4 \dots $.\footnote{To be consistent with our 
treatment of $SU(N)$ below, we use a convention in which 
the spin is labeled by integers rather than half-integers.}  
The space of spherical harmonics is precisely 
reproduced by ground states in the boundary theory 
$(\alpha, \id)$ when we send $\a$ (and hence $k$) to infinity. 
For finite $\a$, the angular momentum $j$ is cut off at a 
finite value $j = \min(2\a, 2k-2\a) \leq 2 \a$. This means
that the brane's world-volume is `fuzzy' since resolving 
small distances would require large angular momenta. The
relation between branes on $SU(2)$ and the familiar non-%
commutative fuzzy 2-spheres \cite{Hop,Mad} was fully analyzed 
in \cite{AlReSc1} and it provides the only known example
of a open string non-commutative geometry beyond the 
familiar case of branes in flat space \cite{DoHu,ChHo,Vol}. 
The analysis of \cite{AlReSc1} goes much beyond the study
of partition functions as it employs detailed information
on the operator product expansions of open string vertex
operators based on \cite{Run}. Using the results in 
\cite{FFFS,FFFStop} it is easy to generalize all these remarks 
on ordinary conjugacy classes to other groups (see also 
\cite{Hop} for more details and explicit formulas on 
fuzzy conjugacy classes).  

Twisted conjugacy classes are more difficult to understand. 
This is related to the fact that they are never `small'. More
precisely, it is not possible to fit a generic twisted conjugacy 
class into an arbitrarily small neighborhood of the group identity 
unless the twist $\omega$ is trivial. This implies that the 
spectrum of angular momenta in $Z_{\a\a}^\om$ is not cut off 
before it reaches the obvious large momentum cut-off that is 
set by the volume of the group, i.e.\ by the level $k$. For
large $\a$ (and large $k$) the ground states in the boundary 
theory $\Xi = (\a,\omega)$ span the space of functions on the 
generic twisted conjugacy classes $C^\om_u$ \cite{FFFS}. The 
non-commutative geometry associated with twisted conjugacy 
classes with finite $\a$, however, remains to be investigated.    

\subsection{Supersymmetric WZW models} 

Throughout this paper we shall address supersymmetric
WZW-models. The main effect on our considerations shows
up in a shift of the level $k$. In fact, a supersymmetric
$SU(N)$ model at level $K = k+N$ describes strings moving 
on $SU(N)$ with $K$ units of NSNS-flux. The model contains 
currents $J^a$ satisfying the relations of a level $k+N$ 
affine Kac-Moody algebra along with a multiplet of free 
fermionic fields $\psi^a$ in the adjoint representation 
of $su(N)$. It is well known that one can introduce new 
{\em bosonic currents}  
$$ 
J^a_\B \ := \ J^a + \frac{i}{k} f^{a}_{\ bc} \psi^b \psi^c 
$$ 
which obey again the commutation relations of a current 
algebra but now the level is shifted to $ K-N = k$. The 
fermionic fields $\psi^a$ commute with the new currents. 
This means that the theory splits into a product of a 
level $k$ WZW model and a theory of three free fermionic 
fields. 

This split is consistent with the boundary conditions we 
study. We want to impose gluing conditions $J^a(z) = 
\Lambda \bar J^a(\bz)$ and $\psi^a (z) = \pm \Lambda \bar 
\psi^a(\bz)$ along the boundary $z = \bz$. This implies 
$J^a_\B (z) = \Lambda \bar J^a_\B (\bz)$ since $\Lambda$ 
is an automorphism of the Kac-Moody algebra so that, in 
particular, its action on the product of the fermionic 
fields is intertwined by the structure constants of the
Lie algebra. Hence, boundary conditions of the level $K$ 
model are described by boundary conditions for the free 
fermions and for a level $k$ bosonic current. Since the
fermionic and the bosonic sectors decouple, we can 
restrict our attention to the latter. But eventually
our results should be interpreted in terms of the 
original supersymmetric background which carries 
$K = k+N$ units of NSNS-flux.     
  
\section{Brane dynamics on group manifolds} 

The central goal now is to understand the dynamics of branes
on group manifolds deep in the stringy regime when the group 
manifolds become small in string units. As a starting point 
it is useful to consider condensation processes in a large
volume expansion, i.e.\ an expansion in powers of $1/k$. There 
one can make reliable and complete statements on the 
renormalization group flows using rather elementary techniques. 
As usual, the stringy regime is much more difficult to 
attack so that we cannot claim to have a complete understanding 
of the renormalization group flows and fixed points for small 
group volumes. But fortunately, a large number of condensation 
processes have been looked at in the past. In fact, the questions
we are considering are very closely related to the Kondo problem 
for which a lot of technology has been invented during the last 
decades. As was remarked in \cite{AlReSc2,AlSc2}, the results carry 
over to the study of gauge field condensates on group manifolds.  
Finally, the comparison with the large volume scenario suggests 
that the essential processes are captured by our analysis.

\subsection{Brane dynamics at large volume} 

In our discussion of the large volume dynamics we follow very 
closely the studies of \cite{AlReSc2}. We consider a stack of 
$M$ identical and symmetry preserving branes of type $\Xi = (\om,
\a)$. This configuration preserves the full $\asg_k$ chiral algebra
and hence we find all the $\asg_k$ currents among the boundary 
operators of the corresponding conformal field theory. The 
rest of the field content depends very much on the particular 
brane $\Xi$ we consider. But in a supersymmetric theory after 
removal of the tachyonic modes, these additional fields will 
become more and more massive as we decrease the size of our 
group manifold.\footnote{A large number of boundary fields 
becomes marginal in the limit $k \rightarrow \infty$} Since 
we are ultimately interested in the stringy regime, this 
justifies to concentrate on the currents which are the only 
massless fields away from the $k\rightarrow \infty$ limit. In 
other words, we restrict to perturbations of the form 
\begin{equation} \label{Spert} S_{\rm pert} \ = \ \int_{\pl \Sigma} 
dx A_a J^a(x) 
\end{equation}
where $x$ is the coordinate on the boundary $\pl \Sigma$ 
of the world-sheet and $A_a, a=1, \dots, \dim G$ is a set of
$M\times M$ Chan-Paton matrices. Adding a perturbation of this form
will give the gauge fields and transverse scalars on the
brane a constant vacuum expectation value $A_a$.    
\smallskip

The rules of perturbative string theory relate the effective
action for the fields $A_a$ to the correlation functions of 
the boundary currents $J^a(x)$. But the latter are completely 
determined by the usual Ward identities and hence they are 
entirely independent of the boundary condition $\Xi$ 
we have selected. This observation implies that the associated 
terms in the brane's effective action are universal, i.e.\ they 
can be computed once and for all without reference to the boundary 
condition we are looking at. The results of these computations can 
be copied from \cite{AlReSc2}, 
\begin{equation}
\label{effact}
{\cal S}_{M \Xi} (A)  \ = \ \tr \left( 
    - \frac14 \,[\, A_a \, , \, A_b\, ]\  [\, A^a \, ,\,  A^b\, ]
    + \frac{i}{3k}\, f^{a b c}\, A_a  \, [\, A_b \, , \, A_c\, ]
    + \mbox{\rm const} \, \right)\ \ .
\end{equation}
Here, $f^{abc}$ are the structure constants of the Lie algebra 
$\sg$ and $\tr$ is the trace on the space of $M \times M$-matrices. 
From eq.\ (\ref{effact}) we obtain the following equations of 
motion  
\begin{equation} 
\label{eom} 
\Bigl[\, A^a  \;\, , [\, A_a \, , \,  A_b\, ] \, - \, \frac{i}{k}\, 
 f_{a b c}\, A^c \,\Bigr]  \ = \ 0 \ \ . 
\end{equation} 
As we remarked before, neither the effective action nor the 
equations of motion depend on the brane $\Xi$. But they 
certainly depend on the number $M$ of branes that we stack 
together through the size of the matrices $A_a$. 
\smallskip

Now we have to study solutions of the equations (\ref{eom}). 
It turns out that there are basically two types of solutions.
The first one is given by a set of $\dim G$ pairwise commuting 
$M \times M$ matrices $A_a$. It comes as a $M \cdot \dim G $ 
parameter family of solutions corresponding to the number of eigenvalues
appearing in $\{ A_a\}$. The same kind of solutions appears 
also for branes in flat backgrounds and the interpretation is 
known from \cite{Wit}. They describe individual rigid translations 
of the $M$ branes on the group manifold. Since each 
brane's position is specified by $\dim G$ coordinates, the number of 
parameters matches nicely with the interpretation. Moving 
branes around in the background is a rather trivial operation 
so that we need not  consider this type of solutions any 
further.   

There exists a second type of solutions to eqs.\ (\ref{eom}) 
which is a lot more interesting. In fact, any $M$-dimensional 
representation of the Lie algebra $\sg$ can be used to solve
the equations of motion. At least for untwisted branes, i.e.\ 
for $\omega = \id$, the interpretation of these solutions
was found in \cite{AlReSc2}. Let us describe the answer for 
general $\Xi = (\alpha,\id)$ and an irreducible $M$-dimensional 
representation $\s$ of 
$\sg$. In this case, the stack of branes 
$\a$ decays into a superposition of branes wrapping various
different conjugacy classes. Which branes appear in the 
final configuration is determined by the Clebsch-Gordan
multiplicities $\cN$ of the finite dimensional Lie algebra
$\sg$. More precisely, one finds 
\begin{equation}
\label{kinvcond} 
 M \ (\alpha,\id) \ \longrightarrow \ \sum_\gamma \ 
   \cN_{\s \a}^{\ \gamma} \ (\gamma,\id) \ \  
\end{equation} 
where $M = \dim(\sigma)$. The support for this statement comes
from both the open string sector and the coupling to closed 
strings (see \cite{AlReSc2}). 

A simple check of the rule \eqref{kinvcond} can be performed, 
if we extend the effective action for a stack of branes $\a$ 
by including non-constant gauge fields $A_a$. The extended  
action $\hat S$ has been derived in \cite{AlReSc2} and it 
is given by 
$$ \hat S_{M (\alpha,\id)} (A_a) \ = \ S_{M \dim(\alpha)\, (0,\id)}
   (Y_a + A_a) \ \ . 
$$
Here $A_a \in \Mat(M \dim(\alpha))$ and $Y_a \sim {\bf 1}_M \otimes
y_a$ involves the $\dim(\a)$-dimensional irreducible representation 
$y_a$ of the Lie algebra $\sg$. The construction of $\hat S$ 
is obviously consistent with the decay \eqref{kinvcond} which 
implies that $M \dim(\alpha)$ branes of the type $(0,\id)$ can 
decay into $M$ branes of type $(\alpha,\id)$. Moreover, 
the formula for $\hat S$ has been derived within string perturbation 
theory in \cite{AlReSc2}. There $\hat S$ was identified as a special
linear combination of Yang-Mills and Chern-Simons theory on 
a fuzzy sphere depending on $\a$. Yang-Mills theories on 
fuzzy spheres are discussed in \cite{FFT,Wat,Madbook}. The 
Chern-Simons term is considered in \cite{Klim}. For branes
on $S^3$, the action $\hat S$ appears instead of the 
non-commutative Yang-Mills theory which was derived in 
\cite{SeiWit} to describe the dynamics of branes in flat 
space with a non-vanishing B-field.   

Irreducible $M$ dimensional representations $\sigma$ of $\sg$ are still 
stationary points of this extended action $\hat S$. To test the rule 
\eqref{kinvcond} we study arbitrary fluctuations $\delta A_a$ of the 
gauge field $A_a 
= \Lambda_a + \delta A_a  \in \Mat(M \dim(\a))$ around a stationary 
point $\Lambda_a \sim \Lambda_a \otimes {\bf 1}_{\dim(\a)}$. Here 
$\Lambda_a$ are the representation matrices of the $M = 
\dim(\s)$-dimensional irreducible representation $\s$ of $\sg$. By 
construction, $\Lambda_a$ and $Y_a$ commute so that their sum 
$\Lambda_a + Y_a$ gives the tensor product representation $\sigma 
\times \alpha$ which decomposes into a sum of irreducibles $\gamma$
with multiplicities $\cN_{\s\a}^{\ \gamma}$. Comparison with the 
construction of $\hat S$ shows that 
$$ \hat S_{M (\alpha,\id)} (\Lambda_a + \delta A_a) \ = \ 
   \sum_\gamma \hat S_{\cN_{\s\a}^{\ \gamma} \, (\gamma,\id)} 
   (\delta A_a) \ + \ \dots  
$$ 
where we omitted terms involving massive fields that come with
open strings stretching between different branes. The resulting 
action for the fluctuation field $\delta A_a$ contains all the 
terms that are predicted by the rule \eqref{kinvcond}. 

Let us finally remark that the final configuration on the right 
hand side of \eqref{kinvcond} is only metastable. Whenever a 
superposition of branes appears in the final state one can 
find a renormalization group flow into another configuration 
of branes with lower mass. These flows, however, are generated
by non-constant gauge fields $A_a$. For detailed explanations
and computations the reader is referred to \cite{AlReSc2}. 

\subsection{Condensation in the stringy regime} 

Now we would like to understand the dynamics of branes in the stringy 
regime. Proceeding  along the lines of the 
previous subsection would force us to include all the higher 
order corrections to the effective action. Unfortunately, this 
problem is even more complicated than finding the non-abelian
Born-Infeld action. Hence, we cannot hope to get a complete 
picture of the brane dynamics in the stringy regime. 

But we could be somewhat less ambitious and ask whether the 
solutions we found in the large volume limit possess a 
deformation into the small volume theory and if so, which 
fixed points they correspond to. In this way we may overlook 
new stationary points of the stringy effective action that 
have no well behaved large $k$ limit. On the other hand, the
reduced program has a positive and very beautiful solution
that is known from the work on the Kondo effect.

The Kondo model is designed to understand the effect of 
magnetic impurities on the low temperature conductance 
properties of a conductor. The latter may have electrons
in several conduction bands. Let us say that there are 
$k$ such bands. Now we can build several currents from 
the basic fermionic fields. Among them is the spin current
$\vec{J}(y)$ which gives rise to a $\asg_k$ 
current algebra. The coordinate $y$ measures the radial 
distance from a spin $s$  impurity at $y=0$ to which 
the spin current couples. This coupling involves a 
$2 s + 1$-dimensional irreducible representation 
$\vec{\Lambda} = (\Lambda_a, a=1,2,3)$ of $su(2)$ and
it is of the form
\begin{equation} \label{Kondo} 
    H_{\rm pert} \ = \ \lambda \Lambda_a J^a(0) \ \ . 
\end{equation}
The operator $H_{\rm pert}$ acts on the tensor product 
$V^\sigma \otimes \cH$ of the Hilbert space $\cH$ for 
the unperturbed theory with the $2 s + 1$ -dimensional 
quantum mechanical state space of our impurity. The 
formula (\ref{Kondo}) is simply the Hamiltonian 
formulation of the perturbations we would like 
to study, as one can see by comparison with formula 
(\ref{Spert}) above. 

Fortunately, a lot of techniques have been developed 
to deal with perturbations of the form (\ref{Kondo}). In 
fact, this problem is what Wilson's renormalization 
group techniques were designed for. From the old
analysis we know that there are two different cases
to be distinguished. When $2 s > k$ 
(`under-screening') the low temperature fixed point 
of the Kondo model appears only at infinite values
of $\lambda$. On the other hand, the fixed point 
is reached at a finite value $\lambda = \lambda^*$ of 
the renormalized coupling constant $\lambda$ if $ 2 s 
\leq k$ (exact- or over-screening resp.). In the latter 
case, the fixed points are described by non-trivial
(interacting) conformal field theories. We can 
summarize the results on the spectrum of the fixed
points by the formula \cite{AffLud} 
\begin{equation} \label{charpert}  
  \tr_{V^\sigma \o \cH_j}\left( q^{H_0 + 
  H_{\rm pert}}\right)_{\lambda = \lambda^*}^{\rm ren} 
  := \sum_l N_{\sigma j}^{\ l} \chi_l(q)\ \ .  
\end{equation} 
Here, $H_0 = L_0 + c/24$ is the unperturbed Hamiltonian, 
the superscript $\ ^{\rm ren}$ stands for `renormalized'
and $V^\s$ denotes the representation space of the 
representation $\sigma$ of $su(2)$ or, more generally, 
of an arbitrary simple Lie algebra $\sg$. The space $\cH_j$ 
can be any of the $\asg_k$-irreducible subspaces in the 
physical state space $\cH$ of the theory. Formula \eqref{charpert} 
means that our perturbation with some irreducible representation 
$\sigma$ interpolates continuously between a building block 
$\dim(\sigma) \chi_j(q)$ of the partition function of the 
UV-fixed point (i.e.\ $\lambda = 0$) and the sum of characters 
on the right hand side of the  previous formula, 
\begin{equation} \label{abs} M \,  \chi_j (q) \ 
   \longrightarrow \ \sum_l N_{\sigma j}
    ^{\ l } \chi_l(q)\ \ , 
\end{equation}
where $M = \dim(\sigma)$. We will now use this rule to find the 
spectra for the decay product of a stack of $M$ branes $\Xi$. To this
end, we need to start from the partition function describing 
open strings stretching between a stack of $M$ branes 
of the type $(\a,\omega)$ and a single brane $(\b,\om)$.
This is given by $M$ times the partition function 
(\ref{part}), i.e.\ by a sum of characters with coefficients
being integer multiples of $M$. We can now employ 
our rule (\ref{abs}) to determine the partition function of 
the system after perturbation with some irreducible 
$M$-dimensional representation $\sigma$ of $\sg$. The 
result is
\beqa Z^\om_{M\a \, \b }(q)  & := &  M Z^\om_{\a\, \b} (q) 
\ \longrightarrow \   \sum_{j \in \cJ_k} {\bfn}_{j \a}
    ^{\om;  \b} \sum_l N_{\sigma j}^{\ l} \chi_l (q) \nn \\[2mm] 
& = &  \sum_{\c \in \cJ^\om_k} \ n_{\sigma \a}^{\om; \c} \ 
 Z^\om _{\c\, \b}(q)
\ \ . \nn 
\eeqa
Here we used the property (\ref{nfus}) of the coefficients 
$n$ to express the right hand side as a linear combination 
of known partition functions. Since the coefficients on the 
right hand side are independent of $\beta$, we can summarize 
the result of our simple computation by the rule
\begin{equation}
\label{condensation}
  M\,  (\a,\omega) \ \longrightarrow \ \sum_{\c\in \cJ^{\om}_k} \ 
  n_{\sigma \a}^{\om; \c}  \, (\c,\omega) \ \ 
\end{equation}
without any reference to the spectator brane $(\beta,\om)$. Here,
$M$ denotes the dimension of the representation $\sigma$
of the finite dimensional Lie algebra $\sg$ and $\a \in 
\cJ^\om_k$. This is the kind of result that we were 
looking for. Let us note that this reproduces the 
picture that we sketched at the end of the previous 
subsection when $\omega = \id$. In this case, the 
numbers $n$ specialize to the fusion rules of $\asg_k$
and they approach the fusion rules of the Lie algebra
$\sg$ when we send $k$ to infinity so that we recover 
the formula \eqref{kinvcond}. The formula \eqref{charpert}
is the content of the `absorption of the boundary spin'-principle 
\cite{AffLud} and it was previously applied to investigations
of brane dynamics in \cite{AlSc2,BarKon}.

\section{Conserved charges and twisted K-theory} 
We would like to see whether the described brane dynamics obey some 
conservation laws, i.\,e.\ if we can assign charges to the branes that are
conserved in physical processes. So we are looking for some discrete 
abelian group $C(X)$, where $X$ denotes the physical background, and a map
from arbitrary brane configurations to $C(X)$ such that the map is invariant 
under renormalization group flows.
\smallskip

Let us denote the charge of a brane $(\alpha, \omega)$ by $q_{(\alpha,\om)}
\in C(X)$. From the process \eqref{condensation} where a stack of 
$\dim(\sigma)$ branes of type $(\alpha,\omega)$ condenses we get  
\begin{equation}
  \label{ccc}
  \dim(\sigma) \, q_{(\alpha,\omega)}\ =\ 
   \sum_{\gamma\in \cJ^{\om}_{k} } \ \bfn^{\om; \gamma}_{\sigma\alpha}\, q_{(\gamma,\om)}\ \ . 
\end{equation}
These equations express the condition for charge conservation under
all the processes we have identified in the previous section. The 
requirement that eqs.\ \eqref{ccc} possess solutions places strong 
constraints on the group $C(SU(N),K)$ of charges. We shall evaluate them 
completely for untwisted branes in the first subsection. Then we turn 
to the twisted branes which are more difficult to control. Nevertheless, 
we will obtain detailed information on $C(SU(N),K)$. This is then summarized 
in the last subsection and compared to what is known on the twisted 
K-groups $K_H^*(SU(N))$.    

\subsection{The charge of untwisted branes on $\boldsymbol{SU (N)}$}

For branes wrapping ordinary conjugacy classes, i.e.\ $\om = \id$, the
integers $\bfn$ are given by the fusion rules $N$. The evaluation
of eqs.\ \eqref{ccc} for $\alpha = 0$ leads to 
\begin{equation}
q_{\beta}\ =\ \dim(\beta)\,  q_{0}
\end{equation}
for all branes $\beta = (\beta,\id)$ and with $q_0 = q_{(0,\id)}$ being 
the charge of the point-like brane at the group unit $e$. Hence, all the 
brane charges of untwisted branes are integer multiples of $q_{0}$. If
we normalize the charge of the point-like brane by $q_{0}=1$ we 
arrive at  
\begin{equation} \label{qbeta}
q_{\beta}\ =\ \dim(\beta) \ \ \ .
\end{equation}
These equations form a subset of the equations \eqref{ccc}. But we 
can see that the charges $q_\b = \dim(\b)$ solve the full set of 
eqs.\ \eqref{ccc} in the limit $k\to\infty$ where the $N$ are just 
the Clebsch-Gordan multiplicities of the simple Lie algebra $su(N)$. 
In this limit, the equations express that the dimension of a 
tensor product of $su(N)$ representations is a sum of dimensions
of its irreducible subrepresentations. For finite $k$, however,
the fusion rules $N$ differ from the Clebsch-Gordan multiplicities
of $su(N)$ so that typically the right hand side of eqs.\ \eqref{ccc} 
with $q_\b = \dim \b$ is smaller than the left hand side.  Hence, the
equations can only hold, if they are evaluated modulo some integer $x$
that we need to determine. Charges then take values in the group $\QZ_x$.
\medskip

Let us first look at a simple example, $X=SU(2)$. For level $k$ the labels 
$\alpha$ lie in the range $0,\dots,k$. We have a geometrical understanding
of what the possible D-branes are. They are given by conjugacy classes
which form 2-spheres embedded in $SU(2)\cong S^3$. Their radius depends 
on $\alpha$ and for $\alpha=0,k$ they degenerate to a point. $\alpha=0$ 
describes a D0-brane at the origin $e$, $\alpha=k$ a D0-brane at $-e$.
\smallskip

\begin{figure}
\begin{center}
\scalebox{0.8}[0.2]{\input{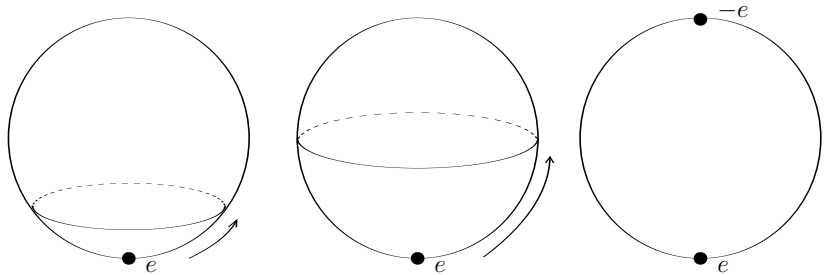}}
\caption{\label{su2branen} Brane dynamics on $S^3$: 
A stack of D0-branes at $e$ can decay to 
a D2-brane. Putting more and more D0-branes at $e$ the resulting brane will
be localized further and further away from the group unit and eventually 
the decay product will be a single D0-brane at $-e$.}
\end{center}
\end{figure}

Now consider a stack of D0-branes at $e$. This stack is expected to decay
into a D2-brane on a 3-sphere with finite volume. If we put more and more 
D0-branes together the radius of the resulting D2-brane will first grow, 
then decrease, and finally a stack of $k+1$ D0-branes will decay to a 
D0-brane at $-e$ (see fig.~\ref{su2branen}).  If we assign charge~$1$
to the D0-brane at $e$ and want the charge to be conserved, the D0-brane 
at $-e$ must have charge $k+1$. On the other hand we could just translate 
the D0-brane from $e$ to $-e$, and by taking orientation into account, 
this would lead to charge $-1$. Thus we have to identify $k+1$ and $-1$ 
which means that charge is only well-defined modulo $k+2$.

We can obtain the same result in a more algebraic way. To this end, we 
evaluate \eqref{ccc} for the simple current $\sigma=J=k$ and the fundamental 
2-dimensional representation $\alpha = 1$. This gives
\begin{equation}
  (k+1)\cdot 2 \ = \  \dim(k) \cdot 2 \ = \ q_{k-1} \ = \ k \ \ \ ,
\label{modkp2}
\end{equation}
where we used that the product of the simple current with the fundamental
representation of $\asg_k$ gives the unique representation with label 
$\beta = k-1$ and $\dim(\beta) = k$. The equation \eqref{modkp2} can only 
hold modulo $x=k+2$. One can show that this choice of $x$ is consistent 
with all processes, i.e.\ that 
$$  \dim(\s) \, \dim(\a) \ = \ \sum_\b \ N^{\ \b}_{\s \a}\, \dim(\b) \ \mod \ 
    (k+2) \ \ . $$   
As we shall see later, it is always sufficient to evaluate the charge 
conservation condition only for simple currents and  fundamental 
representations. The resulting restrictions are strong enough to 
guarantee charge conservation for all processes.  
\medskip

We are now turning to the more general case of $X=SU(N)$. The task is to
find the largest number $x$ such that \eqref{ccc} is fulfilled modulo $x$.
As we can generate all representations out of the fundamental ones, 
$\omega_i$, $i=1,\dots, N-1$, we can reduce our problem to processes involving
stacks of $\omega_i$-branes. In other words, the general charge conservation
condition is fulfilled if
\begin{equation}
  \dim(\sigma) \, q_{\omega_i}\ =\ \sum_{\beta}\ N^{\ \beta}_{\sigma\omega_i}
  \, q_{\beta} \ \mod \  x \end{equation} 
for all $i = 1, \dots, N-1$ and  $\sigma \in \cJ_k$. A rigorous prove 
of this statement can be found in Appendix \ref{appB}.
\smallskip

Denote by $J=k\omega_1$ the generator of the simple current group $\QZ_N$ of 
$\widehat{su}(N)_k$. It can be shown that it suffices to evaluate the 
equations \eqref{ccc} for stacks of $\dim(J)$ fundamental branes (see 
appendix~\ref{appB}). Thus, the charge conservation condition reduces to
\begin{equation}
  \dim(J) \, q_{\omega_i} \ =\ \sum_{\beta}\ N^{\ \beta}_{J\omega_i}\, 
  q_{\beta} \ \mod \ x 
\end{equation} for all $i = 1, \dots, N-1$.  
Taking the difference between both sides with $q_\a = \dim(\a)$ 
inserted, gives the following $N-1$ numbers $a_i$,  
(see \eqref{ai})
\begin{equation}
\label{aiuntwisted}
  a_i \:=\ \dim(J) \, \dim(\om_i) \ - \ \sum_{\beta}\ N^{\ \beta}_{J\omega_i}\, 
  \dim(\b )\ = \ \frac{(k+1)\dots(\widehat{k+i})\dots(k+N)}{(i-1)!\,(N-i)!}
  \ \ \ 
\end{equation}
where the hat over a factor indicates that this factor is omitted.
These numbers have to vanish modulo $x$. This means that $x$ is given by 
the greatest common divisor of these numbers. It can be shown (see appendix 
\ref{appC}) that $x = \gcd(a_i)$ is given by 
\begin{equation}
\label{x}
  x \ =\ \frac{k+N}{\gcd(k+N,\lcm(1,\dots,N-1))} \ \ \ .
\end{equation}
Hence, the charge group of the untwisted branes for $X = SU(N)$  
is $\QZ_x$ with $x$ as in formula \eqref{x}.  

\subsection{Charges of twisted branes on $\boldsymbol{SU (N)}$} 

Let us now take a look at branes that wrap twisted conjugacy classes. As 
gluing automorphism we choose the reflection $\omega$ of the Dynkin diagram. 
Their action on the vertices of the Dynkin diagram induces the following
 map on the weight space,
\begin{equation}
  \varpi (\lambda_1,\dots,\lambda_{N-1})\ =\ (\lambda_{N-1},\dots,\lambda_1)\;,
\end{equation}
where the $\lambda_i$ are (finite) Dynkin labels. Details on our notations
and some fundamental results on the representation theory of $su(N)$ can 
be found in Appendix \ref{appA}. 
\smallskip

As in the untwisted case we get a charge conservation condition,
\begin{equation} \label{twchar} 
  \dim(\sigma) \, q_{\alpha}
\ = \ \sum_{\b \in \cJ^\om_k}\ \bfn^{\om; \beta}_{\sigma \alpha }\, q_{\beta}
\ \ \ \mbox{ for all } \ \ \a \in \cJ^\om_k\ \ .  
\end{equation}
We would like to perform a similar analysis as in the untwisted case but
we are faced with the problem that the integers $\bfn = \bfn^\om$ are a 
lot more difficult to handle than the fusion rules $N$. But even though 
we are not able to fully exploit these conditions, we can get some severe 
constraints on the charge group by symmetry considerations.
\smallskip

We will see that the numbers $\bfn$ are invariant under the action of
some simple currents~$I$,
\begin{equation} \label{invI}
  \bfn^{\ \beta}_{I\!\sigma\, \alpha}
 \ =\ \bfn^{\ \beta}_{ \sigma\, \alpha} \ \ \ .
\end{equation}
To derive this result we look at the explicit expressions \eqref{annuluscf} 
for the numbers~$\bfn$ and investigate what happens under the action of a 
simple current~$I$:
\begin{equation*}
\begin{split}
  \bfn^{\ \beta}_{I\!\sigma\, \alpha}
&=\ \sum_{\lambda \in \cJ^\omega_k}\ 
\frac{S^{\omega \,*}_{\lambda \beta}
S^{\omega}_{\lambda \alpha}
S_{\lambda\,I\!\sigma}}{S_{\lambda 0}}\\[2mm]
&=\ \sum_{\lambda \in \cJ^\omega_k}\ 
e^{2\pi i Q_I(\lambda)}\ \frac{S^{\omega \,*}_{\lambda \beta}
S^{\omega}_{\lambda \alpha }S_{\lambda\sigma}}{S_{\lambda 0}}\ \ \ .
\end{split}
\end{equation*}
Here $Q_I(\lambda)$  is the monodromy charge of $\lambda$ with respect
to the simple current $I$. If it is zero, we infer that the coefficients 
$\bfn$ are invariant under the action of the simple current.

For a symmetric weight $\lambda = \varpi\lambda \in \cJ^\omega_k$ we know 
that
\begin{equation}
  \varpi (J^i\, \lambda)\ =\ J^{N-i}\, \lambda \ \ \ .
\end{equation}
This implies immediately that $Q_J(\lambda)=Q_{J^{N-1}}(\lambda)$. If $N$ 
is odd, it follows that $Q_{J^i}(\lambda)=0$ for all $i= 1, \dots, N-1$. 
If $N$ is even, we can only deduce that $Q_{J^i}(\lambda)=0$ for $i$ even.
We thus arrive at the result that 
\begin{equation}
  \bfn^{\ \beta}_{J^i\!\sigma\, \alpha} \ =\ \bfn^{\ \beta}_{\sigma\, \alpha} 
\end{equation}
for arbitrary $i$ if $N$ is odd and for even $i$ if $N$ is even which 
is the precise formulation of the invariance properties of $\bfn$ we 
anticipated in eq.\ \eqref{invI}. 
\smallskip
 
Assuming that there is at least one twisted brane which can be 
assigned a charge with value $1$ we immediately deduce the 
following condition on the unknown integer $x_\omega$ 
\begin{equation}
  \label{dimcon}
  \dim(J^i\!\sigma)\ =\ \dim(\sigma)\ \mod \ x_\omega \ \ , 
\end{equation}
where the values for $i$ depend on whether $N$ is even or odd, 
as formulated before. 

Let us first concentrate on the case that $N$ is odd. 
Using \eqref{dimcon} with $\sigma= 0,\omega_i$ we obtain
\begin{align}
    \mspace{180mu}\dim(J)&\ =\ 1  &&   \mod \  x_\omega
\mspace{180mu} \\[2mm]
    \dim(J\omega_i)&\ =\ \dim(\omega_i)  &&  \mod \ x_\omega\ \ \ . 
\end{align}
The two relations combine into the following statement for the 
numbers $a_i$ that were defined in eqs.\ \eqref{aiuntwisted} 
above, 
\begin{equation}
  a_i\ \ =\ \ \dim(J)\, \dim(\omega_i) - \dim(J\omega_i) \ = \ 0  
\ \mod \ x_\omega\ \ . 
\end{equation}
By definition, the greatest common divisor of these numbers $a_i$ is $x$ 
and hence we deduce that $x_{\omega}$ is a divisor of $x$, i.e.\ that the 
order of an element in the charge group for twisted branes cannot exceed 
the order of the charge subgroup from untwisted branes. It can be shown 
that $x_\omega = x$ does imply eqs.\ \eqref{dimcon} but we cannot exclude 
that the eqs.\ \eqref{twchar} force $x_\omega$ to be smaller than $x$.

If $N$ is even, we find $x_{\omega}$ not so strongly restricted by the 
eqs.\ \eqref{dimcon}. Introducing the integers $b_0 = \dim(J^2)-1$ and 
$b_i = \dim(J^2 \omega_i) - \dim (\omega_i)$ for $i=1, \dots,N-1$, 
one can show that $x_\omega$ must divide $\gcd(b_i)$. Note that 
$\gcd(b_i)$ is a possibly non-trivial integer multiple of $x$. For 
$SU(4)$ we still get the result that $x_{\omega}$ divides $x$ but 
already for $SU(6)$ one finds situations where \eqref{dimcon} can 
be fulfilled modulo $x_{\omega} > x$. There is some evidence that 
eqs.\ \eqref{dimcon} provide enough restrictions for $N=0\mod 4$ to 
guarantee that $x_\omega$ divides $x$.  

\subsection{Comparison with twisted K-theory} 

Before we explain what is known about the twisted K-groups 
$K_H^*(SU(N))$, let us briefly summarize the results for $C(SU(N),K)$
that we obtained in the previous two subsections. The charge group 
that governs the dynamics of branes in a $\widehat{su}(N)_k$ WZW-model 
is
\begin{equation} \label{NSUNK}
C(SU(N),K) \ = \ \QZ_x \oplus \bigoplus_{\nu = 1}^s  \QZ_{x_\nu}
\end{equation} 
where $x$ is given by \eqref{x} and $x_\nu, \nu = 1, \dots,s$ divide
$x_\omega$. In case $N \leq 5$ we know that $x_{\omega}$ must divide 
$x$ and this remains true for $N>5$ as long as $N$ is odd. For even 
$N \geq 6$ we can only show that $x_\omega$ divides $\gcd(b_i)$ with 
the integers $b_i$ being introduced in the last paragraph of the 
previous subsection. In general, $\gcd(b_i)$ could be some possibly 
non-trivial integer multiple of $x$ but it is very likely that 
$\gcd(b_i) = x$ when $N=0 \mod 4$. Furthermore, we have some hints 
that $x$ and $x_{\omega}$ are equal for $N=3$ from direct calculations 
of the numbers $\bfn$ with small values of $k$. 

As we have seen above, branes wrapping ordinary conjugacy classes can 
all be obtained from stacks of point-like branes. This guarantees that 
there is a unique way to assign charges to such branes as we have 
seen in our discussion leading to eqs.\ \eqref{qbeta}. Hence, untwisted
branes contribute a single cyclic subgroup to the group of charges
$C(SU(N),K)$. For branes wrapping twisted conjugacy classes, similar 
arguments do not exist. As a consequence, we cannot exclude the 
existence of several independent charge assignments for twisted 
branes. The summation over $\nu = 1, \dots, s$ in eq.\ \eqref{NSUNK}
reflects this fact. Actually, it seems to be rather likely that 
$s >1$ for $N > 3$.  
\medskip

According to the proposal of Bouwknegt and Mathai, our results
on $C(SU(N),K)$ should be compared to the twisted K-groups $K_H^*(SU(N))$. 
Unfortunately, the latter have not been computed yet. 
\smallskip

The definition of $K_H^*(X)$ uses the space of sections in a bundle 
over $X$ with fiber being the algebra of compact operators on a 
separable Hilbert space. This space of sections can be turned 
into an algebra and it is known that algebras of this form are
classified by elements of $H^3(X,\bZ)$. In other words, there 
exists some way of assigning an algebra $\cA_H$ to any choice of 
$H \in H^3(X,\QZ)$. The K-groups of this algebra is denoted by 
$K^*_H(X)$. If $H$ vanishes the algebra $\cA_H$ factorizes 
globally into functions on $X$ and compact operators. Hence, by 
Morita invariance of K-theory, $K^*_{H=0}(X)$ coincide with 
ordinary K-groups. 
\smallskip

One way to calculate such K-groups makes use of 
Atiyah-Hirzebruch spectral sequences. These start 
from the de Rham cohomology groups and then proceed 
through a sequence of complexes whose cohomology 
stabilizes after a finite number of steps. The 
resulting cohomology provides some information on 
the desired K-group, though there is still some 
extension problem to solve. Generically, the latter 
may have several solutions. In any case, the problem 
of these computations for $K_H^*(G)$ starts earlier 
because almost nothing is known about the differentials
that appear in the sequence of complexes. Only for the
first non-trivial step, the required differential was 
obtained by Rosenberg in \cite{Ros}. This suffices to 
compute the twisted K-group for $G = SU(2)$. The 
result is 
$$ K_H^*(SU(2)) \ = \ Z_K \ \ . $$
Here $H = K \Omega_3$ and $\Omega_3$ is the 
normalized volume form of the unit sphere. For 
$G = SU(3)$, Rosenberg's results still allow to 
show that 
$$ K_H^*(SU(3)) \ = \ \QZ_r + \QZ_r\ \ ,$$
where $r$ is known to divide $K$. If all the 
higher differentials that are not determined by
the result of Rosenberg would vanish, then one
would get $r = K$. Hence, the comparison with 
our CFT results suggests that the higher 
differentials do not vanish, at least for 
even $K = k+3$.    
\smallskip 

It would be highly desirable to get more 
results on the twisted K-groups. At the 
moment, the restrictions on $C(SU(N),K)$ 
that we obtained by studying renormalization 
group flows in WZW-models provide highly 
non-trivial predictions for $K^*_H(SU(N))$.
It was suggested to us by Wassermann that 
the techniques in \cite{RosSch,Was} could 
lead to a computation of $K^*_H(SU(N))$
which would employ the results of \cite{FrHoTe} 
on equivariant twisted K-theory.

\section{Conclusions and open problems}

In this work we studied brane dynamics on $SU(N)$ and formulated
conservation laws for these dynamics. The conserved charges 
take values in some finite abelian group $C(SU(N),K)$. While
we were not able to determine $C(SU(N),K)$ completely, we 
obtained a number of strong restrictions on its structure. 
These are reflected in our formula \eqref{NSUNK}. Our results
are consistent with the proposal $C(SU(N),K) = K^*_H(SU(N))$ 
of \cite{BouMat}, but since so little is known about $K^*_H$ 
the comparison was restricted to $N=2,3$.  
\smallskip

The main difficulties in our analysis of $C(SU(N),K)$ were 
related with branes wrapping twisted conjugacy classes. A 
better understanding of the numbers $n^{\omega}$ that 
determine the partition functions for such theories would
certainly lead to more detailed information on the group
of charges. This applies, in particular, to the study of
$SU(N)$ with $N$ even and to the question whether twisted
branes can support several independent charge assignments. 
It would be interesting to re-interprete the numbers 
$n^{\om,\b}_{\s\a}$ for finite $\a,\b$ within the framework 
of non-commutative geometry. This is possible for $\omega
= \id$ and in this case it leads to fuzzy geometries. A 
similar interpretation for branes wrapping twisted 
conjugacy classes does not exist. 
\smallskip
   
An extension of our discussion to other groups $G$ is possible. 
Most of the basic ideas we have used do not depend on the 
specific choice $G = SU(N)$. Only in our evaluation of the 
charge conservation condition \eqref{ccc} in Section 4 we
exploited some simplifying features that hold for $G=SU(N)$. 

It would also be interesting to go beyond these examples and
to understand the importance of twisted K-theory for branes
in curved backgrounds more generally from the nature of the
fields that condense upon bound state formation. Such arguments 
would be analogous to the relation between ordinary K-theory and 
tachyon condensation (see e.g.\ \cite{Wit1,HaMo} and references 
therein). Note, however, that in the cases we studied above 
the dynamics is driven by massless fields rather than 
conventional tachyons. Let us also stress that all the 
processes we have considered involve finite stacks of 
branes. This is is some contrast to the construction of 
twisted K-theory for non-torsion H-fields which involves 
taking some limit $M \rightarrow \infty$ \cite{BouMat}. 
This has motivated the authors of \cite{BouMat} to 
speculate about some relation with processes on an 
infinite stack of branes.  
\smallskip

Finally, we would like to mention that many of the results
on branes in WZW-models descend to other models of conformal
field theory through orbifold and coset constructions. 
Thereby, our results could be used to extend the 
investigations in \cite{ReRoSc} and they should bear some 
relevance even for the behavior of branes in Gepner models 
which describe strings and branes on certain Calabi-Yau spaces 
deep in the stringy regime. We plan to return to these issues in 
a forthcoming publication. 
\bigskip
\bigskip

\noindent
{\bf Acknowledgements:} We would like to thank A.Yu.\ Alekseev, 
V.\ Braun, I.\ Brunner, A.\ Carey, C.S.\ Chu, D.\ Freed, R.\ 
Helling, V.\ Mathai, G.\ Moore, A.\ Recknagel, T.\ Quella and 
A.\ Wassermann for stimulating and very useful discussions. This 
research was carried out in part for the Clay Mathematics Institute.

\begin{appendix}
\section{Some representation theory}
\label{appA}
In this appendix we will briefly review some facts in 
$\widehat{su}(N)$-representation theory. Details can be found e.\,g.\ in
\cite{DiFr}.

An affine weight $\lambda$ can be expanded in fundamental weights,
\begin{equation*}
\lambda=\lambda_0\omega^0+\lambda_1\omega^1+\dots+\lambda_{N-1}\omega^{N-1} \;.
\end{equation*}
The expansion coefficients are the Dynkin labels. When we consider 
representations at level $k$, the zeroth Dynkin label is fixed
by the others,
\begin{equation*}
\lambda_0=k-\sum_{i=1}^{N-1}\lambda_i \;,
\end{equation*}
therefore $\lambda$ is determined by its finite Dynkin labels 
$(\lambda_1,\dots,\lambda_{N-1})$.

The fundamental weights are then given by
\begin{equation*}
\omega_i=(0,\dots,0,\begin{array}[t]{c}1\\i\end{array},0,\dots,0)\;,
\end{equation*}
the vacuum representation is $(0,\dots,0)$.

We are interested in integrable highest-weight representations. We find 
that their highest weight $\lambda$ has to be dominant, i.\;e.\ the Dynkin
labels of $\lambda$ have to be non negative integers. For a given level $k$
there are only finitely many dominant weights restricted by
 \begin{equation*}
\sum_{i=1}^{N-1}\lambda_i\le k \;.
\end{equation*}

Instead of using Dynkin labels we can specify a weight $\lambda$ in terms
of its partition
\begin{equation*}
\lambda=\{\ell_1;\ell_2;\cdots;\ell_{N-1}\}
\end{equation*}
where
\begin{equation*}
\ell_i=\lambda_i+\cdots+\lambda_{N-1} \;.
\end{equation*}

The dimension of the representation of the finite simple Lie algebra $su(N)$
belonging to the highest weight $\lambda$ can be easily given by
the partition,
\begin{equation}
\label{dimfor}
\dim(\lambda)=\prod_{1\le i<j\le N}\frac{\ell_i-\ell_j+j-i}{j-i} \;,
\end{equation}
where $\ell_N=0$.

Partitions are also useful in calculating Clebsch-Gordan
coefficients via the Littlewood-Richardson rule (see e.\,g.\cite{DiFr}).
The fusion rules of the affine Lie algebra can be obtained from the
Clebsch-Gordan coefficients by a suitable truncation.

Let us consider an example that we will need for our discussion. We consider 
the fusion of the simple current generator $J=(k,0,\ldots,0)$ with a
fundamental weight $\omega_i$. In the tensor product decomposition
we find two representations, $J+\omega_i$ and $J-\omega_1+\omega_{i+1}$ 
(setting $\omega_N=0$). The first one has $\ell_1=k+1$ and is ignored
because of the truncation at level $k$,
the second one remains.

The dimensions of the corresponding representations of the finite Lie algebra 
fulfil
\begin{equation*}
\dim(J)\, \dim(\omega_i)\ =\ \sum_{\beta}\tilde{N}^{\beta}_{J\omega_i}
\dim(\beta)
\end{equation*}
where $\tilde{N}$ denote the finite tensor-product coefficients.

When we substitute $\tilde{N}$ by the fusion rules $N$ of the 
affine Lie algebra,
this equation is not longer valid. In our example, the difference
between both sides is then given by $\dim(J+\omega_i)$ which is 
(using \eqref{dimfor})
\begin{equation}
\label{ai}
a_i:=\dim(J+\omega_i)=
\frac{(k+1)\dots(\widehat{k+i})\dots(k+N)}{(i-1)!\,(N-i)!}\;.
\end{equation}

\section{Some lemmas used in Section 4}
\label{appB}
We consider the affine Lie algebra $\widehat{su}(N)_k$. We denote the 
fundamental weights by $\omega_i$, $i=1,\dots,N-1$. By $q_{\alpha}=
\dim(\alpha)$ we denote the dimension of the irreducible highest-weight 
representation of the horizontal subalgebra corresponding to $\alpha$.

\begin{lem}
Suppose
\begin{align*}
  q_{\sigma}\, q_{\omega_i}&\ =\ \sum_{\beta}\ N^{\ \beta}_{\sigma\omega_i}
 \ q_{\beta} \ \mod x \ \quad 
  \forall \ i,\sigma \ \ .\\
\intertext{Then}
  q_{\sigma}\, q_{\alpha}&\ =\ \sum_{\beta}\ 
 N^{\ \beta}_{\sigma\alpha}\ q_{\beta}\ \mod x \ \quad 
  \forall \ \alpha,\sigma \ \ .
\end{align*}
\end{lem}
\begin{proof}
We will proof the lemma by induction over the sum of the finite Dynkin labels
$\ell_1(\alpha)=\sum_{i=1}^{N-1}\alpha_i$. The equation obviously holds for 
$\ell_1(\alpha)=0$ and for $\ell_1(\alpha)=1$ (fundamental weights).

Suppose now that the assertion is valid for labels with $\ell_1\le \ell$. 
For a label $\alpha$ with $l_1(\a ) \leq l+1$ we denote by
$i=i(\alpha)$ the number between $0$ and $N-1$ satisfying
$\ell_j(\alpha)=\ell+1$ for $1\le j\le i$ and $\ell_j(\alpha)\le \ell$ 
for $j>i$. Clearly the equation holds for weights satisfying $i=0$. By 
induction we show that it holds for all $i$ and therefore for all 
weights with $\ell_1\le \ell+1$.

Let $\alpha$ be a weight with $\ell_1(\alpha)=\ell+1$. Then this weight 
appears once in the fusion of the weight $\alpha'=\alpha-\omega_{i(\alpha)}$ 
with  $\ell_1(\alpha')=\ell$ and the fundamental weight $\omega_{i(\alpha)}$. 
The other weights $\lambda$ appearing in the fusion have $i(\lambda)<i(\alpha)$. 
Assuming that the equation is valid for these $\lambda$ and using the 
associativity of the fusion product we show that the equation holds for 
$\alpha$, 
\begin{equation*}
\begin{split}
\sum_{\beta}\ N_{\sigma\alpha}^{\ \beta}\ q_{\beta}\quad 
&=\quad \sum_{\beta}\ N_{\alpha'\omega_i}^{\ \alpha}
\, N_{\sigma\alpha}^{\ \beta}\ q_{\beta}\\
&=\quad\sum_{\beta}\ \Bigl[\ \sum_{\lambda}\ N_{\alpha'\omega_i}^{\ \lambda}
\, N_{\sigma\lambda}^{\ \beta} - \sum_{\lambda\not= \alpha}
\ N_{\alpha'\omega_i}^{\ \lambda}\, N_{\sigma\lambda}^{\ \beta} \ \Bigr]\ q_{\beta}\\
&=\quad\sum_{\beta}\ \Bigl[\ \sum_{\lambda}\ N_{\sigma\omega_i}^{\ \lambda}
\, N_{\alpha'\lambda}^{\ \beta} - \sum_{\lambda\not= \alpha}\ 
N_{\alpha'\omega_i}^{\ \lambda}\, N_{\sigma\lambda}^{\ \beta} \ \Bigr]\ 
q_{\beta}\\
&\stackrel{\makebox[0pt]{$\scriptstyle \!\mod x$}}{=} 
\quad \sum_{\lambda}\ N_{\sigma\omega_i}^{\ \lambda}\ q_{\alpha'}\, q_{\lambda}
-\sum_{\lambda\not= \alpha}\ 
N_{\alpha'\omega_i}^{\ \lambda}\ q_{\sigma}\, q_{\lambda}\\
&\stackrel{\makebox[0pt]{$\scriptstyle \!\mod x$}}{=} 
\quad
q_{\sigma}\, q_{\omega_i}\, q_{\alpha'}-q_{\alpha'}\, q_{\omega_i}\, q_{\sigma}
+N_{\alpha'\omega_i}^{\ \alpha}\ q_{\s}\, q_{\a}\\
&=\quad q_{\s}\, q_{\a} \ \ .
\end{split}
\end{equation*}
This completes the proof of Lemma 1. 
\end{proof}

We will show in the following that it is sufficient to evaluate the
charge conservation condition for fundamental representations $\omega_i$ and
the simple current generator $J$.
\begin{lem}
Suppose
\begin{align}
  \label{vorausstzg}
  q_{J}\, q_{\omega_i}&\ =\ \sum_{\beta}\ N^{\ \beta}_{J\omega_i}\ q_{\beta} \ \mod x 
   \quad  \forall \ i\ \ . \\
\intertext{Then}
  q_{\sigma}\, q_{\omega_i}&\ =\ \sum_{\beta}\ N^{\ \beta}_{\sigma\omega_i}\ q_{\beta}
\ \mod x \quad 
  \forall \ i,\sigma \ \ .
\end{align}
\end{lem}
\begin{proof}
Let us first remark that the equation certainly holds for $\ell_1(\sigma)<k$, 
because then the fusion matrices $N$ coincide with the finite tensor-product
coefficients.
We are now going to proof the statement:

For all $i=1,\dots,N-1$ and $\ell=0,\dots,k-1$ the following is true:
\begin{itemize}
\item[A] 
\begin{align*}
\sum_{\beta}\ N^{\ \beta}_{\sigma\omega_j}\ q_{\beta}\quad
 &\stackrel{\makebox[0pt]{$\scriptstyle \!\mod x$}}{=} 
\quad q_{\sigma} \, q_{\omega_j}
&\forall \ j=1,\dots,N-1 &\\
& & \forall \ \sigma \text{ with } \ell_1(\sigma)=k \;,\;
&\ell_j(\sigma)\le \ell+1 
\text{ for } j\ge 2 \\
& & & \ell_j(\alpha)\le \ell \text{ for } j\ge i+1 
\end{align*}
\item[B]
\begin{align*}
\dim(\beta)\;\;
 \stackrel{\makebox[0pt]{$\scriptstyle \!\mod x$}}{=} 
\quad 0 \quad \forall\ \beta \ \ \text{ with }\ \  \ell_1(\beta)=k+1\;,\;&
\ell_j(\beta)\le \ell+2 \text{ for } j\ge 2\\
& \ell_j(\beta)\le \ell+1 \text{ for } j\ge i+1 \ \ . 
\end{align*}
\end{itemize}

We proof this proposition by induction over $\ell$ and $i$. We start with
$\ell=0,i=1$. Part~A is fulfilled because of \eqref{vorausstzg}. For part 
B consider a weight 
$\beta$ with $\ell_1(\beta)=k+1, \ell_2(\beta)\le 1$. Then
$\beta=J+\omega_j$ for some $j$. This is just the truncated weight in
the fusion of $J$ and $\omega_j$, therefore
\begin{equation*}
\dim(J+\omega_j)\ =\ q_{\omega_j}\, q_J-\sum_{\beta}\ N^{\ \beta}_{\omega_j J}
\  q_{\beta} \quad
 \stackrel{\makebox[0pt]{$\scriptstyle \!\mod x$}}{=} \quad 0 \;.
\end{equation*}

We note that the statements A and B for $\ell,i=N-1$ 
are equivalent to the statements for $\ell+1,i=1$.
For the induction process we only have to show the step 
$(\ell,i)\Rightarrow(\ell,i+1)$.

Assume that $A_{\ell,i}$ and $B_{\ell,i}$ are valid. Let $\alpha$ be a label
with $\ell_1(\alpha)=k,\ell_2(\alpha)\le \ell+1$. The fusion of $\alpha$
and $\omega_i$ differs from the finite tensor-product decomposition
just by representations $\beta$ with 
$\ell_1(\beta)=k+1,\ell_2(\beta)\le \ell+2$ and $\ell_{i+1}(\beta)\le\ell+1$.
From $B_{\ell,i}$ we know that their dimensions vanish modulo $x$ and hence
\begin{equation}
\sum_{\beta}N^{\ \beta}_{\alpha\omega_i}\, q_{\beta}\quad
 \stackrel{\makebox[0pt]{$\scriptstyle \!\mod x$}}{=} 
\quad q_{\alpha} \, q_{\omega_i} \quad 
\text{for $\,\ell_1(\alpha)=k\, ,\ \  \ell_2(\alpha)\le\ell +1$} \;.
\end{equation}

Now we will proof $A_{\ell,i+1}$. Let $\alpha$ be a 
label with $\ell_1(\alpha)=k$
and $\ell_2(\alpha)=\cdots=\ell_{i+1}(\alpha)=\ell+1$,
$\ell_j(\alpha)\le \ell$ for $j\ge i+2$. We then define
\begin{equation*}
\alpha '\ =\ \{k;\underbrace{\ell;\cdots;\ell}_{i-\text{times}};\ell_{i+2};\cdots\}
\;.
\end{equation*}
$\alpha$ occurs once in the fusion of $\alpha '$ and $\omega_i$, all the
other labels occurring in the fusion fulfil the requirements of $A_{\ell,i}$.
Hence
\begin{equation*}
\begin{split}
\sum_{\beta}\ N_{\a \omega_j }^{\ \beta}\ q_{\beta} 
&\ =\quad \sum_{\beta}\ N_{\alpha'\omega_i}^{\ \alpha}
\, N_{\a \omega_j }^{\ \beta}\ q_{\beta}\\
&=\quad\sum_{\beta}\ \bigl[\ \sum_{\lambda}\ N_{\alpha'\omega_i}^{\ \lambda}
\, N_{\lambda \omega_j }^{\ \beta} - \sum_{\lambda\not= \alpha}\ 
N_{\alpha'\omega_i}^{\ \lambda}\, N_{\lambda \omega_j}^{\ \beta} \ 
\bigr]\ q_{\beta}\\
&\stackrel{\makebox[0pt]{$\scriptstyle \!\mod x$}}{=} 
\quad
\sum_{\lambda,\beta}\ N_{\a' \omega_j}^{\ \lambda}
\, N_{\lambda \omega_i}^{\ \beta}\ q_{\beta} - \sum_{\lambda\not= \alpha}
\ N_{\alpha'\omega_i}^{\ \lambda}\  q_{\lambda} \, q_{\omega_j} \\
&\stackrel{\makebox[0pt]{$\scriptstyle \!\mod x$}}{=} 
\quad \sum_{\lambda}\ N_{\a' \omega_j}^{\ \lambda}\ q_\lambda \, q_{\omega_i}
-\sum_{\lambda}\ 
N_{\alpha'\omega_i}^{\ \lambda}\ q_{\omega_j}\, q_{\lambda}+
N_{\a' \omega_i}^{\ \alpha}\ q_\a \, q_{\omega_j} \\
&\stackrel{\makebox[0pt]{$\scriptstyle \!\mod x$}}{=} 
\quad
q_{\a'} \, q_{\omega_j} \, q_{\omega_i}- q_{\a'} \, q_{\omega_i}\, 
q_{\omega_j} +q_\a \, q_{\omega_j}\\
&=\quad q_\a \, q_{\omega_j} \ \ .
\end{split}
\end{equation*}

Now we have to show $B_{\ell,i+1}$. Let $\beta_0$ be a label  of the form 
\begin{equation*}
\beta_0\ =\ \{ k+1;\underbrace{\ell+2;\cdots;\ell+2}_{i\text{-times}};
\underbrace{\ell+1;\cdots;\ell+1}_{(j-i-1)\text{-times}}
;\ell_{j+1}(\beta_0);\cdots;\ell_{N-1}(\beta_0) \} 
\end{equation*}
with $\ell_{j+1}(\beta_0)\le \ell$ and define 
\begin{equation*}
\beta'=\{k;\underbrace{\ell+1;\cdots;\ell+1}_{i\text{-times}};
\underbrace{\ell;\cdots;\ell}_{\makebox[0pt]{$\scriptstyle (j-i-1)\text{-times}$}}
;\ell_{j+1}(\beta_0);\cdots;\ell_{N-1}(\beta_0)\} \;.
\end{equation*}
Then $\beta_0$ appears once in the finite tensor product 
of $\beta'$ and $\omega_{j}$. It belongs to the representations that are
truncated by going over to the fusion rules of the affine Lie algebra.
For the other truncated representations $\beta$ we know from $B_{\ell,i}$ that
$\dim(\beta)=0 \mod x$. But since $A_{\ell,i+1}$ is applicable to
$\alpha=\beta'$ we get $\dim(\beta_0)=0\mod x$. This completes the 
proof. \end{proof}

\section{Evaluation of $\boldsymbol{\gcd(a_i)}$}
\label{appC}
\begin{lem} Let the numbers $a_i$ be defined as in \eqref{aiuntwisted}. 
Then their greatest common divisor is given by 
\begin{equation*}
  x \ := \ \gcd(a_i) \ =\ \frac{k+N}{\gcd(k+N,\lcm (1,\dots,N-1))} 
  \ \ \ .
\end{equation*}
\end{lem}
\begin{proof}
We are only going to give a sketch of the proof. Let us rewrite the numbers
$a_i$ by introducing
\begin{equation*}
b(N-1) \ =\ \frac{(N-1)!}{\lcm (1,\dots,N-1)}
\end{equation*}
as
\begin{equation*}
  a_i\ =\ \frac{(k+1)\dots(\widehat{k+i})\dots(k+N-1)}{b(N-1)}
\, \binom{N-1}{i-1}\, \frac{k+N}{\lcm (1,\dots,N-1)}\;.
\end{equation*}
An important observation is that the first factor in $a_i$ is
always an integer. As also the binomial coefficient is an integer,
we can see that $x$ is a divisor of all $a_i$.

It remains to show that it is already the greatest common divisor.
Let $p$ be a prime number. We determine the maximum $y$
and the corresponding $i$ such that $p^y|(k+i)$. Then one can show that 
$p\not| \;\frac{a_i}{x}$. 
\end{proof}

\end{appendix}

\def\gaw{Gawedzki}

\end{document}